\documentclass[conference]{IEEEtran}
\IEEEoverridecommandlockouts

\usepackage{physics}
\usepackage[ruled,nofillcomment,linesnumbered]{algorithm2e}

\usepackage{makecell}
\usepackage{braket}
\usepackage{graphicx}
\usepackage{svg}
\usepackage{amsmath}
\usepackage{amssymb}
\usepackage{caption}
\usepackage{subcaption}
\usepackage{floatrow}
\newfloatcommand{capbtabbox}{table}[][\FBwidth]

\usepackage{fancyhdr}
\usepackage[normalem]{ulem}
\usepackage[hyphens]{url}
\usepackage[sort,nocompress]{cite}
\usepackage[final]{microtype}
\usepackage[bookmarks=true,breaklinks=true,color links,citecolor=blue,linkcolor=blue,urlcolor=blue]{hyperref}

\usepackage[noblocks]{authblk}

\begin{document}

\title{Faulty towers: recovering a functioning quantum random access memory in the presence of defective routers}

\author[1,2]{D.~K.~Weiss}
\author[1,2]{Shifan Xu}
\author[1,2]{Shruti Puri}
\author[2,3]{Yongshan Ding}
\author[1,2]{Steven M.~Girvin}

\affil[1]{Departments of Physics and Applied Physics, Yale University, New Haven, CT 06511, USA}
\affil[2]{Yale Quantum Institute, Yale University, New Haven, CT 06511, USA}
\affil[3]{Department of Computer Science, Yale University, New Haven, CT 06511, USA}
\maketitle

\begin{abstract}
Proposals for quantum random access memory (QRAM) generally have a binary-tree structure, and thus require hardware that is exponential in the depth of the QRAM. For solid-state based devices, a fabrication yield that is less than $100\%$ implies that certain addresses at the bottom of the tree become inaccessible if a router in the unique path to that address is faulty. We discuss how to recover a functioning QRAM in the presence of faulty routers.
We present the \texttt{IterativeRepair} algorithm, which constructs
QRAMs layer by layer until the desired depth is reached. This algorithm utilizes ancilla flag qubits which reroute queries to faulty routers. We present a classical algorithm \texttt{FlagQubitMinimization} that attempts to minimize the required number of such ancilla. For a router failure rate of $1\%$ and a QRAM of depth $n=13$, we expect that on average 430 addresses need repair: we require only 1.5 ancilla flag qubits on average to perform this rerouting.
\end{abstract}

\section{Introduction}

A quantum-random-access-memory (QRAM) device can be utilized to access data in superposition. A wide variety of quantum algorithms require QRAM, including Grover's search \cite{Grover1996}, quantum simulation \cite{Babbush2018} and quantum machine learning \cite{Jaques2023}.  
An extensive introduction to QRAM as well as proposed use cases can be found in Ref.~\cite{Dalzell2023}. Various hardware proposals have been put forth in recent years to realize a QRAM device, based on circuit-quantum acoustodynamics \cite{Hann2019}, surface-acoustic-wave phonons \cite{Wang2024}, superconducting circuits \cite{Weiss2024}, silicon vacancies \cite{Chen2021}, Rydberg atoms \cite{Hong2012} and quantum optics \cite{Giovannetti2008architectures}. 

All of these proposals employ a binary-tree architecture, ensuring that the memory is truly ``random access." Thus, the quantum circuit for querying a QRAM is shallow, requiring time that is only logarithmic in the size of the memory $N$ (linear in the depth $n=\log_{2}(N)$ of the tree) \cite{nielsenchuang2010, Giovannetti2008, Hann2021}. This comes at the cost of requiring a number of quantum elements (qubits, routers, etc.) that scales linearly in $N=2^n$.
Indeed, realizing a binary-tree QRAM with only three or four address qubits requires dozens, if not hundreds, of quantum elements \cite{Weiss2024, Chen2021}. With exponential in $n$ quantum elements, we generally expect exponentially many errors during a QRAM query. Nevertheless, recent work has shown that the infidelity of a query only scales poly-logarithmically in the size $N$ of the memory \cite{Hann2021}. 

Here, we focus not on random errors that occur during a query, but rather on how to mitigate manufacturing faults originating during construction or fabrication of the QRAM tree. This is a well-studied problem in classical computer architectures, solved by fabricating memory cell arrays with spare rows and columns \cite{Kuo1986}. Fabrication defects are a known issue affecting larger-scale solid-state quantum devices, with individual nonfunctioning quantum elements occurring generally at the single-digit percentage level \cite{Arute2019, Zhu2021}. Previous work has explored techniques for mitigating such fabrication defects in the context of surface-code architectures \cite{Auger2017, Debroy2024, Lin2024}. However, these issues are more problematic in the context of a binary-tree architecture, as any faulty quantum routers (either in the body of the tree or at the bottom) render down-stream addresses inaccessible from the root of the tree. 

Previous work by Kim {\it et al.} considers a related problem of faulty quantum memory cells in a QRAM \cite{Kim2023}. However, in QRAM architectures for accessing classical data, the data is generally not stored at the bottom of the tree in quantum memory cells \cite{Weiss2024, Hann2021}. Instead, the classical data is copied into the state of the so-called bus qubit (which has been routed to the bottom of the QRAM tree) by using the classical data to decide whether to flip the state of the bus \cite{Hann2021}. This classical data is thus stored only in the classical controller in the time before the data-copy operation. It is thus more natural and relevant to consider the issue of how faulty quantum routers either at the bottom or in the interior of the QRAM tree affect the ability to successfully perform a QRAM query.

\begin{figure}[t]
    \centering
    \includegraphics[width=\linewidth]{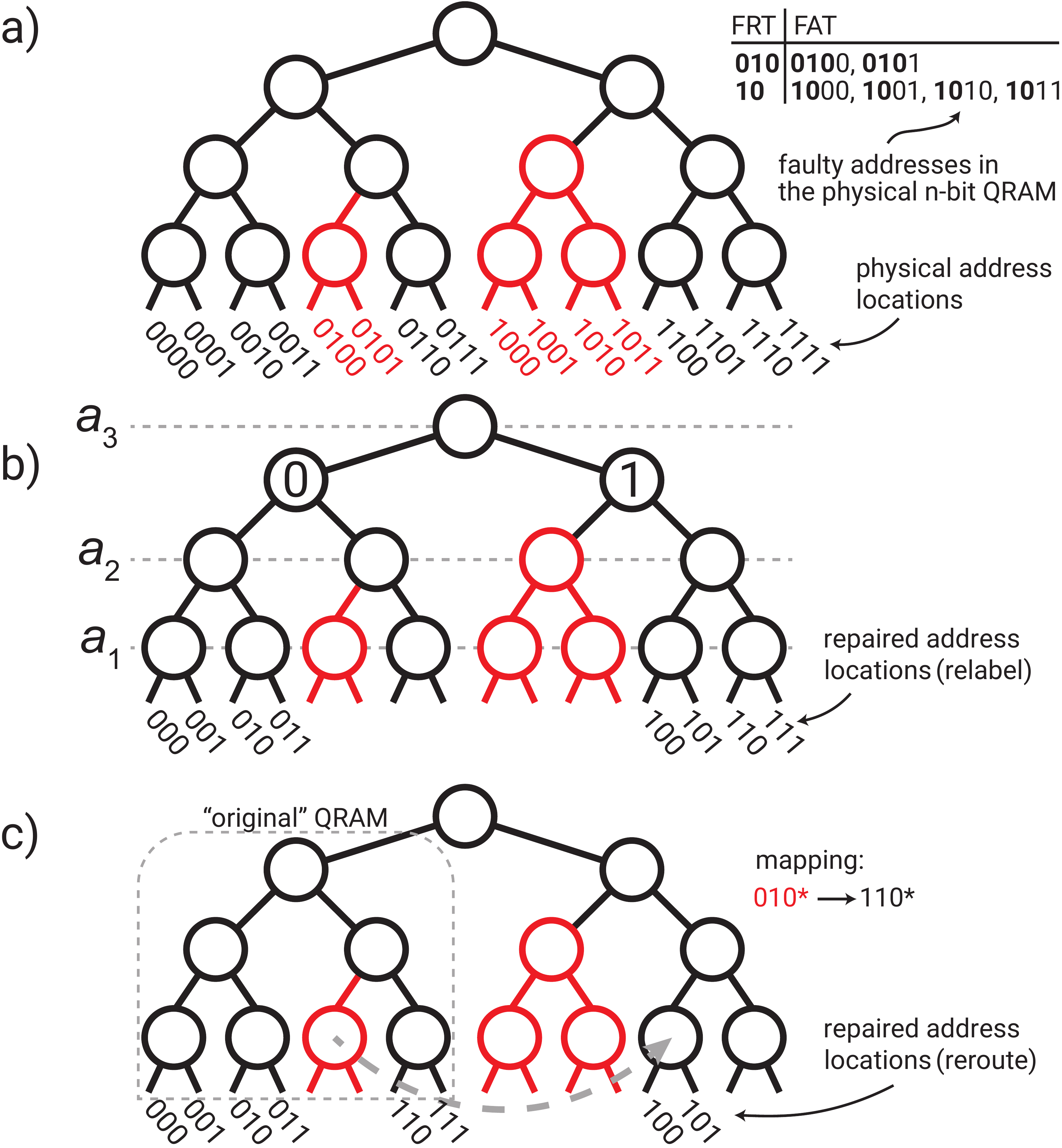}
    \caption{\label{fig:fig1} Schematic of the proposed QRAM repair schemes. In this example, the goal is to recover a functioning 3-bit QRAM from a faulty 4-bit QRAM. (a) We are provided with a faulty-router table (FRT) that specifies which routers or links are faulty, shown in red. The FRT then uniquely specifies which physical addresses are unreachable. (b) In some cases, we can relabel the locations of classical data in the memory. This \texttt{RelabelRepair} requires only three address qubits and sets certain routers to route permanently in a single direction. 
    In the instance shown, it is both routers in the second layer of the tree, and their fixed routing directions are indicted by the inscribed 0 or 1.
    The classical-data locations in the 3-bit QRAM are those of the 4-bit QRAM, with the second-most significant bit selected out by the one-way routers. (c) With the \texttt{IterativeRepair} algorithm, faulty address locations on the repairable side of the tree are mapped to available addresses on the spare side, layer by layer (in this example, we repair only the bottom layer, so there is only a single iteration). Queries to $a_{3}a_{2}a_{1}=10*$ (the user does not have access to $a_{4}$, which is used for repair) which would arrive at a faulty router on the repairable side now instead arrive at the indicated routers on the spare side.
    }
\end{figure}

Specifically, the question we address in this work is how to recover a functioning QRAM of reduced size in the presence of defective quantum elements, see Fig.~\ref{fig:fig1}. Both the routers themselves and the quantum links between them can be faulty. (A faulty link makes the router beneath it unreachable, which is equivalent to that router itself being faulty. Thus, from here on we refer mainly to faulty routers for simplicity.) We assume that we are provided with a faulty-router table (FRT) from which we immediately infer the faulty address table (FAT), which encodes which addresses are inaccessible \footnote{The FRT can be constructed based only on local quantum process tomography, which can be parallelized across the tree.}. 
One possible solution is to expand the connectivity of the tree, allowing for routing around and past faulty routers. However, this likely incurs a significant resource overhead in terms of additional coupling elements. We consider instead recovering a working $(n-1)$-bit QRAM given a faulty $n$-bit QRAM (and given that $2^{n-1}$ addresses remain accessible), see Fig.~\ref{fig:fig1}. 

We perform this repair in two different ways, depending on the complexity of the problem. In the simplest \texttt{RelabelRepair} algorithm, we select $2^{n-1}$ of the available addresses that we label from left to right as the addresses of an $n-1$-bit QRAM, see Fig.~\ref{fig:fig1}(a). This algorithm works by treating some routers as ``one-way streets," which route only in one direction and specifically away from parts of the tree with inaccessible routers. The trick is to recover the binary-tree structure where each router (that is not a one-way street) sees an identical number of addresses below to the left and below to the right.

\sloppy In cases where a QRAM cannot be found using \texttt{RelabelRepair} (but $2^{n-1}$ addresses remain accessible), a more sophisticated rerouting algorithm is required. [Ironically, the rerouting function that maps faulty addresses to working addresses can be efficiently performed if one has access to a (working) QRAM]. The \texttt{IterativeRepair} algorithm proceeds by using the QRAM to repair itself layer by layer, see Fig.~\ref{fig:fig1}(b). Each step of the algorithm uses the upper layers of the QRAM to map faulty routers to available ones in the next layer, allowing us to extend one layer further down the tree and thus increase the QRAM depth by one. In this way, we obtain an iterative construction of a functioning $(n-1)$-bit QRAM. It is important to note that each of these effective QRAMs is part of the same physical device. We assume only that we initially have access to a functioning 2-bit QRAM (i.e., the top three routers are not faulty) and that $2^{n-1}$ addresses are available at the bottom of the tree.

The process of mapping faulty addresses to available ones utilizes ancilla flag qubits that are activated if a query is going to arrive at a faulty router. These flag qubits then conditionally flip the address qubits to perform the rerouting. The number of necessary flag qubits depends both on how the flag-qubits are used and on the classical mapping of faulty addresses to available ones. We first present a pedagogical flag-qubit mask technique, where each flag qubit conditionally flips a specific address qubit. We then show how this technique can be optimized with the \texttt{FlagQubitMinimization} algorithm, by encoding a specific bit-flip pattern in each flag qubit. This reduces the number of ancillae needed as compared to the flag-qubit-mask technique.

Our paper is structured as follows. In Sec.~\ref{sec:background} we provide a short introduction to quantum computing, quantum hardware and QRAM to ensure that the paper is self-contained. We additionally discuss previous work in classical-computing architectures for mitigating fabrication defects. We then analyze in Sec.~\ref{sec:faulty_stats} the expected number of faulty addresses given a finite probability that any router is faulty. In Sec.~\ref{sec:repair} we present the \texttt{RelabelRepair} and \texttt{IterativeRepair} algorithms. We discuss the problem of the classical assignment of faulty routers to available ones in Sec.~\ref{sec:flag}, and the impact these assignments have on the number of flag qubits required to perform the rerouting. We evaluate our numerical simulations in Sec.~\ref{sec:evaluation}, and discuss our results and compare to existing literature in Sec.~\ref{sec:discussion}. Finally, we conclude in Sec.~\ref{sec:conclusion}.

\section{Background}
\label{sec:background}
\subsection{Fundamentals of Quantum Computing and Quantum Hardware}

Quantum computing leverages principles of quantum mechanics to process information in fundamentally new ways. Unlike classical bits, a quantum bit (qubit) can exist in superposition, described by the state $\ket{\psi} = \alpha\ket{0} + \beta\ket{1}$. The coefficients $\alpha$ and $\beta$ are generally complex and satisfy $|\alpha|^2 + |\beta|^2 = 1$, ensuring that the quantum state is normalized. Quantum gates manipulate qubits via unitary transformations, which preserve normalization. Inherently quantum phenomena like entanglement allow the states of multiple qubits to be correlated in a manner not possible for classical bits, enabling a quantum speedup for specific tasks as compared to the best known classical algorithms ~\cite{nielsenchuang2010}.

Many platforms have been proposed as the physical basis for building a quantum computer. These include 
superconducting circuits~\cite{Blais2021}, trapped ions~\cite{trappedions}, neutral atoms~\cite{rydberg, Cong2022}, and nitrogen-vacancy-centers in diamond~\cite{NV}. 
It is beyond the scope of this paper to discuss in detail the various advantages and drawbacks offered by each platform. 
For the purposes of this work, it is enough to note that many of these platforms are susceptible to either fabrication defects \cite{Strikis2023, Arute2019}, qubit loss \cite{Cong2022}, or correlated errors (e.g., from cosmic-ray events) that render part of the quantum chip temporarily unusable \cite{McEwen2022}.

\subsection{Quantum Random Access Memory}
\label{sec:QRAM}
\begin{figure}
    \centering
    \includegraphics[width=\linewidth]{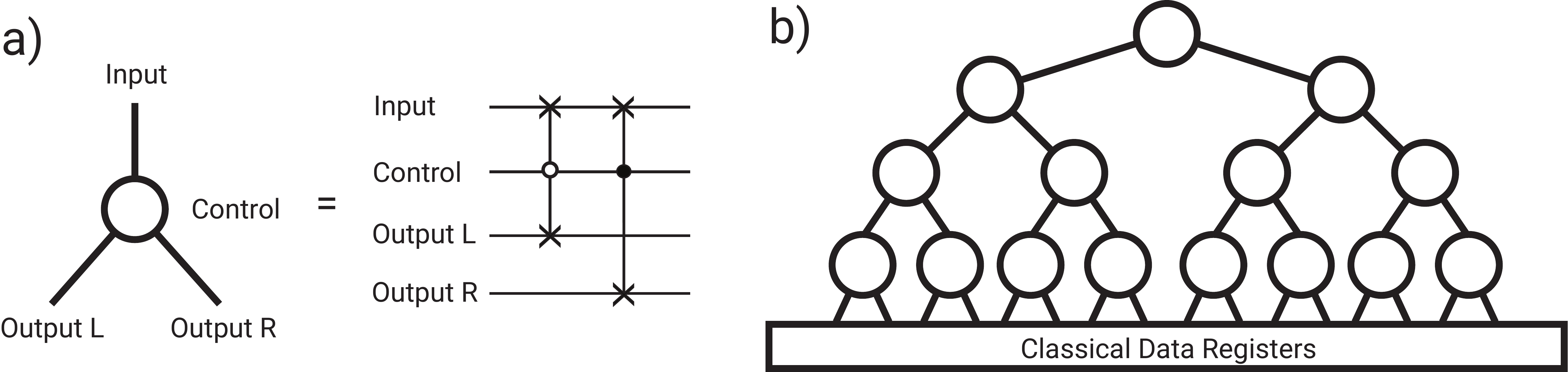}
    \caption{\label{fig:bg} 
    (a) A quantum router consists of four qubits and two CSWAP operations between the control qubit and the input/output data qubits. (b) A tree-structured QRAM consists of recursive quantum routers, with classical data registers located at the bottom of the tree.}
\end{figure}
QRAM is an architecture that enables quantum computers to efficiently access and process large classical datasets into quantum states. This operation is also commonly referred to as a quantum query. Specifically, a quantum query implemented by QRAM consists of performing the unitary operation
\begin{align}
\label{eq:QRAM}
\sum_{i}\alpha_{i}|i\rangle|0\rangle \xrightarrow{\rm QRAM} \sum_{i}\alpha_{i}|i\rangle|d_{i}\rangle,
\end{align}
where the left qubit register represents the state of the address qubits and the right the state of the bus qubit(s). 
The state $|i\rangle$ of the address-qubit register specifies a location $i$ in a classical database storing data $d_{i}$ (where $d_{i}$ could consist of many bits). The goal of the QRAM operation is to perform the data query using a quantum superposition of multiple different addresses. This simultaneously encodes the data corresponding to each address in the state of the bus qubits (we need as many bus qubits as there are bits in $d_{i}$). 

\subsection{Router-based QRAM: Architectures and Operations}
Multiple proposals for the implementation of QRAM have been proposed in recent years~\cite{nielsenchuang2010, Giovannetti2008architectures, Hong2012, Weiss2024, Chen2021, Hann2019, Xu2023, Wang2024}. In this paper, we focus on proposals for physical implementations of a device that could carry out Eq.~\eqref{eq:QRAM} where the fundamental building block is a {\it quantum router}, see Fig.~\ref{fig:bg}(a) \cite{Xu2023, Hann2021}.
Though we use a router-based query architecture for the demonstration of our repair methods, our repairing methodology can be easily extended to other QRAM implementations as well.

The quantum router routes quantum data from an input to one of two different outputs conditioned on the state of a control quantum bit, see Fig.~\ref{fig:bg}(a). Arranging multiple routers in a tree structure generates a router-based QRAM [see Fig.~\ref{fig:bg}(b)], where the two outputs of a parent router serve as the inputs of child routers at the next layer. Hence, quantum data can be routed to multiple different locations in coherent superposition. By leveraging quantum routers, router-based QRAMs can implement $O(\log N)$-latency queries for a memory of capacity $N$ at the cost of $O(N)$ qubits.

By routing address qubits into the tree conditioned on the states of more significant address qubits \cite{Giovannetti2008}, a degree of noise resilience is ensured due to the resulting low entanglement between routers in different branches \cite{Hann2021}. This routing scheme is known as ``bucket-brigade QRAM" \cite{Giovannetti2008}.
After loading in the address qubits, the states of the routers at a given layer of the tree are set by the state of the corresponding address qubit: for example, the most significant address qubit sets the state of the root (topmost) router. The bus qubit(s) are then routed in following the address qubits, reaching locations at the bottom of the tree specified by addresses in the superposition ~\eqref{eq:QRAM}. The classical data is copied into the state of the bus qubit, which is then routed out of the tree, followed by the address qubits. This last ``uncompute" step disentangles the address and bus qubits from the routers in the QRAM tree and completes the operation ~\eqref{eq:QRAM}.  

\subsection{RAM/disk Error and Mitigation: Classical RAM versus QRAM}
Classical RAM and disk error mitigation rely heavily on data redundancy, employing additional hardware to enhance reliability and fault tolerance~\cite{schroeder2009dram}. Techniques such as RAID, introduced in the 1980s, leverage redundant disk drives through mirroring, striping, and parity to guard against disk failures while improving performance~\cite{patterson1988case}. RAM fabrication errors can be solved by a similar technique, which fabricates memory cell arrays with spare rows and columns \cite{Kuo1986}. Later, Error-Correcting Code (ECC) schemes were introduced to embed redundancy directly into RAM chips, using extra bits to detect and correct transient errors during operation~\cite{schroeder2009dram,borucki2008comparison}. By strategically incorporating redundancy, these approaches provide robust error mitigation, forming the backbone of fault-tolerant architectures in classical computing systems.

Dealing with quantum errors similarly relies on redundancy, but the challenges are fundamentally different due to the unique properties of quantum systems. Unlike classical systems, where errors manifest as bit flips or data corruption, quantum errors include decoherence, gate errors, and measurement inaccuracies, which can affect both the amplitude and phase of quantum states~\cite{nielsenchuang2010}. Techniques such as quantum error correction (QEC) use additional qubits—analogous to the redundancy in classical ECC—to encode logical qubits and correct errors without directly measuring the quantum state~\cite{calderbank1996good, Gottesman1998}. Previous work has investigated how QEC might be integrated with QRAM \cite{Jaques2023, HannThesis, Hann2021}. 

Here, as mentioned above, we focus not on correcting quantum errors but instead on how classical faults (fabrication defects, qubit loss, degradation of coherence making certain qubits unusable, etc.) uniquely affect QRAM devices. In traditional quantum systems, such faults 
typically corrupt a single quantum data register. However, QRAM faults arising from failures in an internal router may cause the entire subtree rooted at that router to be inaccessible. This creates a cascading loss of accessibility, rather than isolated data corruption. This unique failure mode demands new error-repair strategies, which motivates this paper describing how to recover a functioning QRAM in the presence of defective routers.

\section{Faulty address statistics}
\label{sec:faulty_stats}

Given a finite and uniform router fabrication failure rate of $\epsilon$, we consider the expected number of faulty addresses $F_{n}(\epsilon)$ at the bottom of an $n$-bit QRAM tree.
This problem is well known under the name of the {\it Galton-Watson branching process} \cite{Harris}, originally studied to model the disappearance of surnames.

\begin{figure}
    \centering
    \includegraphics[width=\columnwidth]{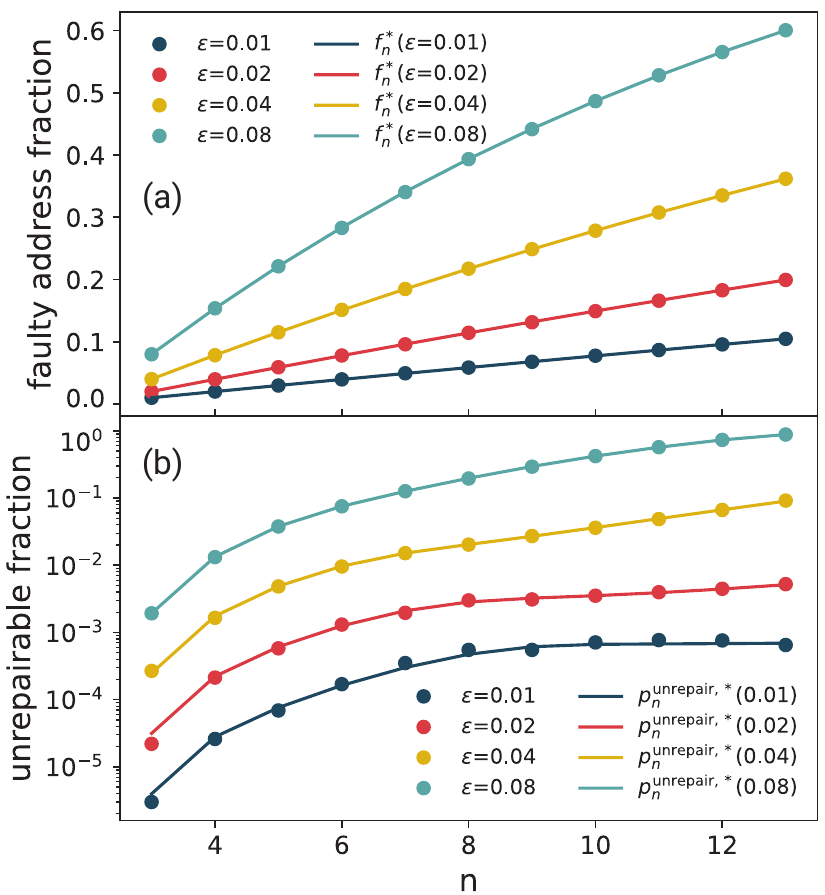}
    \caption{\label{fig:faulty_unrepair}
    Faulty address and unrepairable fractions. For both (a) the average-faulty-address fraction and (b) the fraction of instances that are unrepairable, results obtained from numerics (solid circles) agree excellently with analytics (solid lines). The fraction of QRAM instances that are unrepairable approaches unity once the mean faulty-address fraction approaches and exceeds 0.5. 
    }
\end{figure}

It is most natural to compute the quantity $F_{n}(\epsilon)$ recursively. 
Consider the root (topmost) router, which is connected to every branch of the QRAM. For each branch, if this router fails, it adds 2 faulty addresses. However, it only contributes to the faulty router count if the below routers in the branch are all not faulty. This situation occurs with probability $\epsilon (1-\epsilon)^{n-1}$. There are $2^{n-1}$ such branches, thus the topmost router contributes $2\epsilon(1-\epsilon)^{n-1}2^{n-1}$ to the expected total number of faulty addresses. Continuing down the tree, we may consider each half of the tree as an $n-1$-bit QRAM. Thus, the expected number of faulty addresses is recursively defined as 
\begin{align}
F_{n}(\epsilon) = 2\epsilon(1-\epsilon)^{n-1}2^{n-1} + 2 F_{n-1}(\epsilon).
\end{align}
With the initial condition $F_{1}(\epsilon)=2\epsilon$ (a 1-bit QRAM has two addresses and a single router that can fail with probability $\epsilon$), we obtain
\begin{align}
F_{n}(\epsilon) &= 2^n\epsilon\sum_{k=1}^{n}(1-\epsilon)^{k-1} \\ 
&= 2^{n}(1-(1-\epsilon)^{n}).
\end{align}
In the limit of $\epsilon\ll 1$, we obtain to leading order in $\epsilon$ 
\begin{align}
F_{n}(\epsilon) = \epsilon n 2^n + O(\epsilon^2).
\end{align}
It is useful in the following to define the expected fraction of faulty addresses
\begin{align}
f_{n}(\epsilon) = F_{n}(\epsilon)/2^n = 1 - (1-\epsilon)^n.
\end{align}
In this work, we always take the routers at the top two levels (the three topmost routers) to be not faulty, as any configuration where this is not so is generally unrepairable. We denote the expected number of faulty routers in this case as $F_{n}^{*}(\epsilon)$, which is given by
\begin{align}
F_{n}^{*}(\epsilon) = 4  F_{n-2}(\epsilon).
\end{align}
Thus, the faulty-address fraction is modified to
\begin{align}
f_{n}^{*}(\epsilon) = F_{n}^{*}(\epsilon)/2^{n} =  1 - (1 - \epsilon)^{n-2}.
\end{align}
This expression is in excellent agreement with numerical results, see Fig.~\ref{fig:faulty_unrepair}(a). For a given tree depth $n$ and faulty-router rate $\epsilon$ (keeping the top three routers not faulty), we generate between 100,000 and 1,000,000 tree instances. For each instance we calculate the number of faulty addresses $F_{n}^{*}(\epsilon)$ and the faulty address fraction $f_{n}^{*}(\epsilon)$, and average over all instances, see Fig.~\ref{fig:faulty_unrepair}(a). 

Given the number of faulty addresses $F_{n}^{*}(\epsilon)$, an instance is unrepairable if $F_{n}^{*}(\epsilon)>2^{n-1}$, or alternatively if $f_{n}^{*}(\epsilon)>0.5$. We semi-analytically calculate the unrepairable fraction as follows.
For ease of discussion, we assume here that all routers have the same failure rate $\epsilon$, before returning to the case where we assume the top three routers are non-faulty. It is useful to define the probability $P_{n}(\ell, \epsilon)$ that $\ell$ addresses are available in an $n$-bit QRAM. $P_{n}(\ell, \epsilon)$ can be calculated recursively by summing up two contributions. The first contribution is if the topmost router is faulty, which contributes a probability $\epsilon$ if $\ell=0$. The second contribution then examines the case where the topmost router is not faulty. Then for the $n$-bit QRAM to have $\ell$ addresses available, the two child QRAMs each of depth $n-1$ can have $m$ and $m'$ addresses available respectively, for any (non-negative) $m$ and $m'$ satisfying $\ell=m+m'$. We thus obtain the recursive relationship
\begin{align}
\label{eq:Pnl}
P_{n}(\ell, \epsilon)
&=(1-\epsilon)\sum_{m=0}^{\ell}P_{n-1}(m, \epsilon)P_{n-1}(\ell-m, \epsilon)+\epsilon\delta_{l,0}.
\end{align}
The base case is $n=1,$ with $P_{1}(0, \epsilon) = \epsilon, P_{1}(2, \epsilon)=1-\epsilon$. The probability that a router is unrepairable is then 
\begin{align}
p_{n}^{\rm unrepair}(\epsilon)=\sum_{\ell=0}^{2^{n-1}-1}P_{n}(\ell, \epsilon).
\end{align}
In the case when the routers at the top two levels are never faulty, we need to correct for overcounting three contributions to $p_{n}^{\rm unrepair}(\epsilon)$ where the top three routers could be faulty. The first is the case when the topmost router is faulty, with probability $\epsilon$. The second is the case when the top router is not faulty, but the two routers at the next level are both faulty with probability $\epsilon^2(1-\epsilon)$. The last case is if the topmost router is not faulty, one of the two routers at the second level is faulty, and \emph{at least} one router on the other side is faulty. This occurs with probability $2\epsilon(1-\epsilon)^2[1-(1-\epsilon)^{2(2^{n-2}-1)}]$. Finally, we need to normalize the result to account for the prior assumption that the top three routers are not faulty with probability $(1-\epsilon)^3$. We thus obtain
\begin{align}
p_{n}^{\rm unrepair,*}(\epsilon)&=[p_{n}^{\rm unrepair}(\epsilon)-\epsilon -\epsilon^2(1-\epsilon) \\ \nonumber &\quad -2\epsilon(1-\epsilon)^2(1-\{1-\epsilon\}^{2^{n-1}-2})]/(1-\epsilon)^3,
\end{align}
for the unrepairable probability in the case where the top three routers are assumed to be non-faulty. This expression is in excellent agreement with numerics, see Fig.~\ref{fig:faulty_unrepair}(b).

\section{Repair algorithms}
\label{sec:repair}

\begin{figure}
    \centering
    \includegraphics[width=\columnwidth]{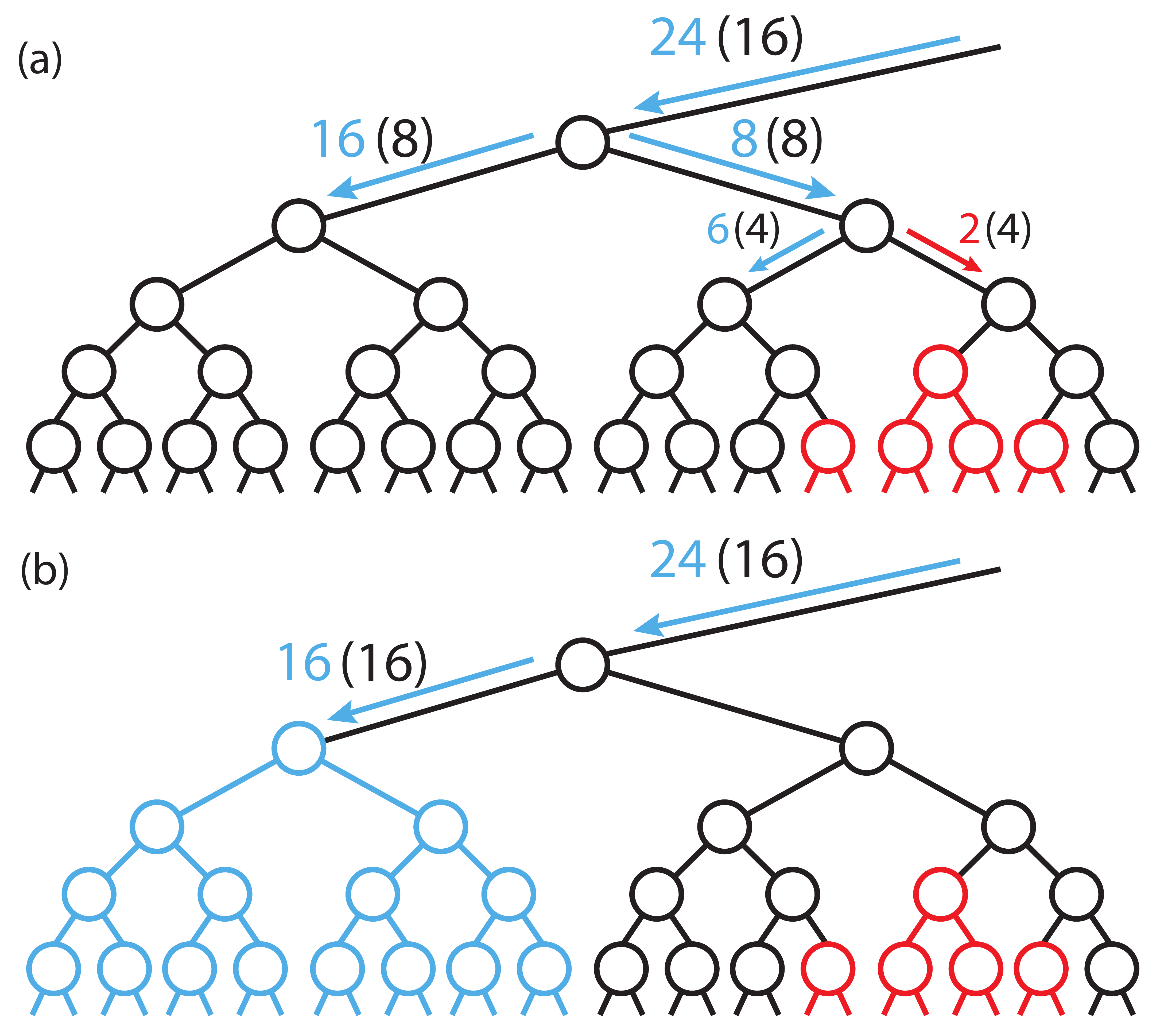}
    \caption{\label{fig:relabel_repair_retreat} Illustration of the \texttt{RelabelRepair} algorithm. We attempt to repair a QRAM by treating certain routers as ``one-way streets." The required number of addresses below each router is shown in parentheses and the actual number of available addresses is indicated in blue (there are sufficiently many addresses) or red (there are insufficiently many). Here, the indicated subtree needs to serve as a 4-bit QRAM. (a) Splitting the topmost router and the child router on the right fails.
    (b) We are forced to retreat, and treat the topmost node as a one-way street to the left. We thus obtain the 4-bit QRAM shown in blue. Of course, if even one router on the left-hand side of the tree is faulty, the algorithm will fail to find an $m=4$ bit QRAM. In such a case, \texttt{IterativeRepair} can be attempted.}
\end{figure}

Given that a depth $n$ QRAM has $2^{n-1}$ addresses available (we say that such a QRAM is {\it repairable}), the question we address here is how to perform a QRAM query on such a faulty device. We describe two algorithms for performing this repair. The first \texttt{RelabelRepair} algorithm is based on the idea that in some cases, not all of the routers in the QRAM tree need to route in both directions, see Fig.~\ref{fig:relabel_repair_retreat}. Of course, most of them must do so in order to properly perform QRAM queries. However, other routers can serve as ``one-way streets," bypassing areas of the QRAM tree containing faulty routers. This algorithm can fail to repair the QRAM even when it is in fact repairable, see Fig.~\ref{fig:relabel}. In such cases, we utilize the \texttt{IterativeRepair} algorithm. This algorithm begins with the largest depth QRAM obtainable using \texttt{RelabelRepair}, and iteratively constructs larger depth QRAMs by rerouting queries from faulty routers to available ones at each layer using smaller depth QRAMs.

\subsection{Overview of the \texttt{RelabelRepair} algorithm}
\label{sec:relabel_repair}

\begin{figure}
    \centering
    \includegraphics[width=\columnwidth]{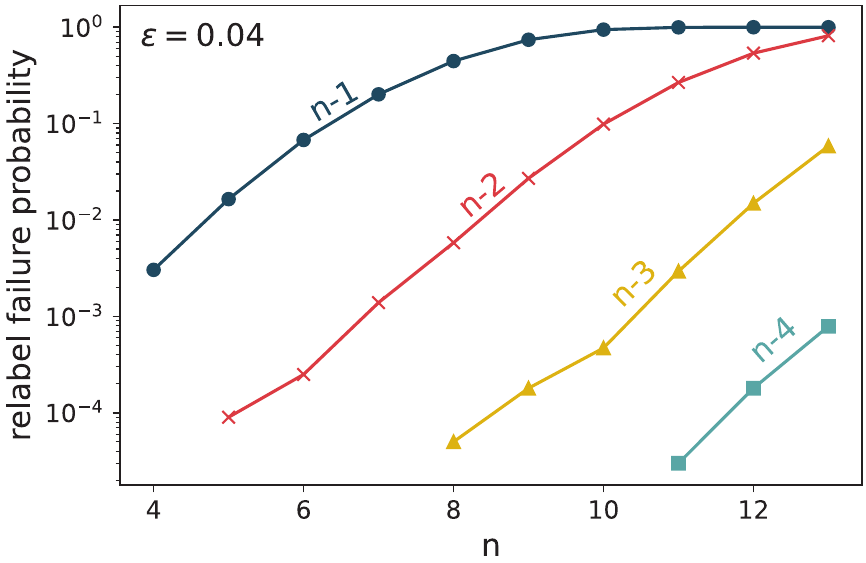}
    \caption{\label{fig:relabel}
    Average failure probability of \texttt{RelabelRepair} as a function of tree depth for $\epsilon=0.04$. This average is only computed among repairable trees.
    As the QRAM depth increases, the probability of \texttt{RelabelRepair} failing for a given desired depth increases exponentially, demonstrating the necessity of \texttt{IterativeRepair}.
}
\end{figure}

\SetCommentSty{textsl}
\begin{algorithm}
\caption{\label{alg:relabel_repair} Algorithm for repairing a QRAM simply by relabelling the addresses. This recursive algorithm assumes the existence of the auxiliary function \texttt{AvailableAddresses} which returns the number of available addresses accessible from the provided router. The algorithm either returns a list of the routers at the bottom of the $m$-bit QRAM or raises an exception if it fails. }
\SetKwFunction{func}{RelabelRepair}
\SetKwProg{Fn}{Function}{:}{}
\Fn{\func{$r$, $m$, $k=1$, $RA=\emptyset$}}{
    \KwIn{router $r$ which contains information about its location in the tree and links to a left child $\ell c$ and right child $rc$ which are themselves routers (or addresses if at the bottom of the tree); desired depth of tree $m$; level of the tree $k$ that Router should serve as}
    \KwOut{list of repaired addresses $RA$}
    \SetKwProg{try}{try}{}{}
    \SetKwProg{catch}{catch}{}{end}
    \SetKwProg{raisee}{raise}{}{}

    \BlankLine
    
    \If{$k==m$\tcp*[r]{we have split $m$ times}}{
        append $r$ location to $RA$\;
        \KwRet{$RA$}
    }
    $a_r\gets$ \texttt{AvailableAddresses}($r.rc$) \;
    $a_\ell\gets$ \texttt{AvailableAddresses}($r.\ell c$) \;
    \If{$a_r \geq 2^{m-k}$ {\bf and} $a_\ell \geq 2^{m-k}$ \tcp*[r]{we can try splitting}}{
        \try{}{
            $RA\gets$ \texttt{RelabelRepair}($r.rc$, $m$, $k+1$, $RA$)\;
            $RA\gets$ \texttt{RelabelRepair}($r.\ell c$, $m$, $k+1$, $RA$)\;
            \KwRet{$RA$}
        }
        \catch(){RelabelRepairFailure}{pass \tcp*[r]{catch splitting failure, try one way}}
    }
    \If{$a_r \geq 2^{m-k+1}$}{
        $RA\gets$ \texttt{RelabelRepair}($r.rc$, $m$, $k$, $RA$)\;
        \KwRet{$RA$}
    }
    \If{$a_\ell \geq 2^{m-k+1}$}{
        $RA\gets$ \texttt{RelabelRepair}($r.\ell c$, $m$, $k$, $RA$)\;
        \KwRet{$RA$}
    }
    \raisee{RelabelRepairFailure \tcp*[r]{Splitting and one way both failed for this router. If above router(s) have been split, this exception can be caught by one-way routing these routers. Otherwise, the algorithm fails.}}{}
}   
\end{algorithm}
We are provided with an $n$-bit QRAM, and we seek an $m$-bit QRAM with $m<n$. (In this work, we usually take $m=n-1$. However, the \texttt{IterativeRepair} algorithm discussed below can make use of QRAMs obtained from \texttt{RelabelRepair} with $m<n-1$.) We are guaranteed that a repairable QRAM has at least $2^m$ available addresses. 
Starting at the root node at layer $k=1$ (the bottom layer is $k=n$), we calculate the number of available addresses on the right and left sides of the QRAM tree, which we denote by $a_{\rm r}$ and $a_{\rm \ell}$ respectively. 
If both $a_{\rm r,\ell} \geq 2^{m-1}$, then it is possible that both the left and the right sides of the tree can host $m-1$ bit QRAMs individually. The topmost router could then serve as a normal two-way router, which we refer to as ``splitting." Alternatively, we could also have either of $a_{\rm r,\ell} \geq 2^{m}$, in which case the topmost router could serve instead as a one-way street to the side where this inequality is satisfied. When a choice between splitting and one-way routing is available, we take a preference for splitting, as only a limited number of routers can serve as one-way streets. We then continue down the tree, keeping track of how many routers along each branch we have successfully split. For instance, in the case where the topmost router is split, we ask if each of its child routers can be split with $2^{m-2}$ addresses available on each side. In the case where the topmost router serves as a one-way street (say to the left), we ask if the child router at the next level on the left can be split into two $m-1$ bit QRAMs. 

The choice between splitting and one-way routing gives the algorithm some flexibility. If it fails at some point further down the tree, we may return to the nearest router where such a choice was made and treat it instead as a one-way router, see Fig.~\ref{fig:relabel_repair_retreat}. Pseudocode for this algorithm is provided in Algorithm~\ref{alg:relabel_repair}. Despite this flexibility, there are cases where for a given $m$ no repaired-repaired QRAM can be found, see Fig.~\ref{fig:relabel}. In such cases an alternative scheme, \texttt{IterativeRepair} (see below), may succeed.

\subsection{Overview of the \texttt{IterativeRepair} algorithm}
\label{sec:iterative}

\begin{figure*}
    \centering
    \includegraphics[width=\columnwidth]{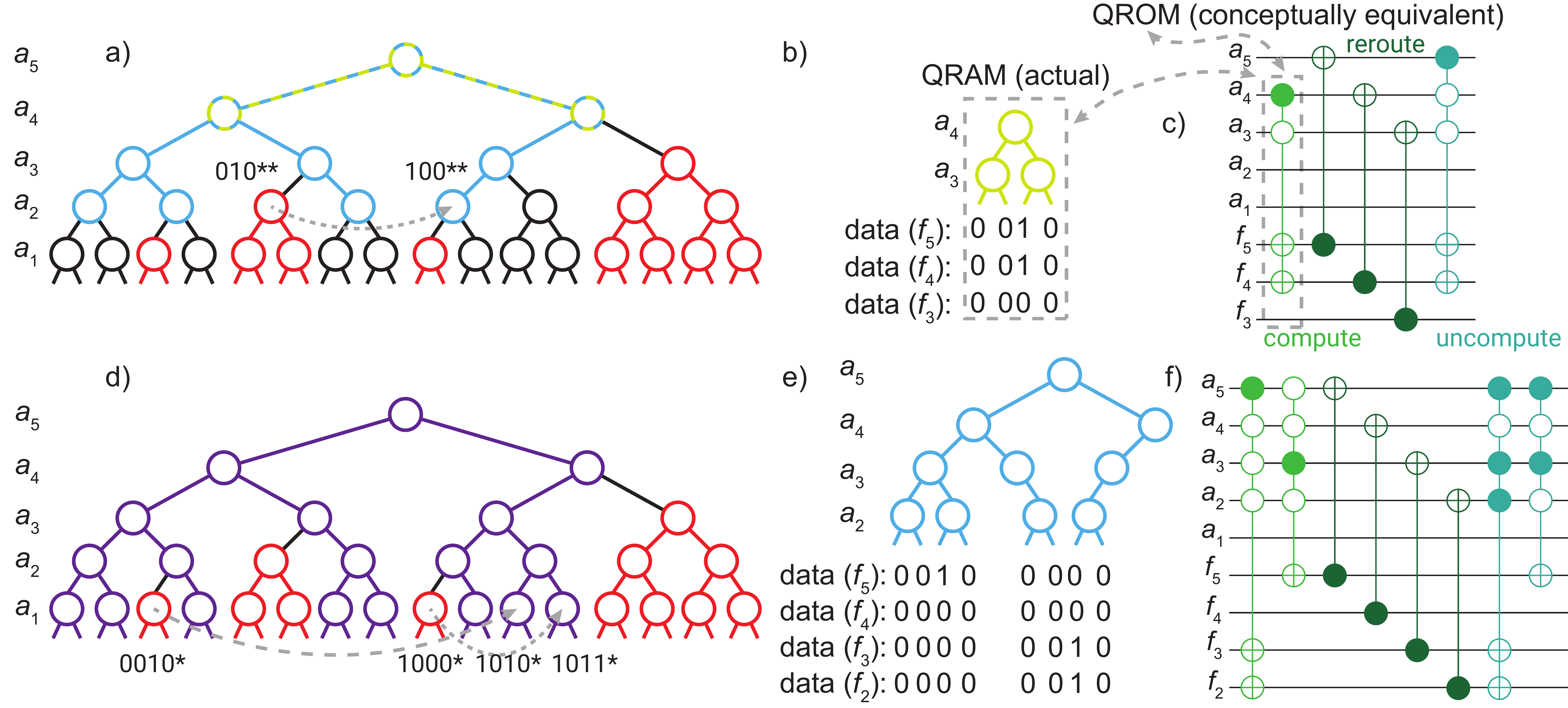}
    \caption{\label{fig:repair_as_you_go} In the \texttt{IterativeRepair} scheme, we construct QRAMs layer by layer, avoiding faulty routers (shown in red). Here, the goal is to obtain a working 4-bit QRAM from the given faulty 5-bit QRAM. (a) 
    The layer second from the bottom of the tree is the first layer needing repair. Repair at each layer proceeds in two steps: first a classical assignment is made of the faulty router to an available router (gray dashed line) followed by (b, c) quantum rerouting of queries that would have arrived at the faulty router. (c) The quantum rerouting is performed using ancilla flag qubit(s) [not shown in (a)] that are activated if a query is to arrive at a faulty router (here, using the flag-qubit mask strategy).
    The gray dashed boxes in (b) and (c) delineate the 2-bit QRAM tree [shown in green in (a)] utilized with associated classical data, along with the conceptually equivalent multiply-controlled Toffoli gate, respectively. 
    The resulting working 3-bit QRAM is shown in blue in (a) and (e). 
    (d) The final layer of the QRAM is repaired with the assignments and reassignments shown in gray. 
    (e, f) The rerouting for this layer is performed using the previously obtained 3-bit QRAM, yielding a working 4-bit QRAM [purple routers in (d)]. 
    }
\end{figure*}

\sloppy In cases where the \texttt{RelabelRepair} algorithm fails, we require a more sophisticated rerouting scheme. Here, we detail the \texttt{IterativeRepair} scheme that repairs the QRAM layer by layer until the desired depth is reached. This scheme works because specifying the location of a faulty router at the $k^{\rm th}$ layer requires only $k-1$ bits (the state the router itself would be set to is irrelevant). Thus, we can iteratively build up larger QRAM trees using smaller ones, assuming only that initially a 2-bit QRAM is available and $2^{n-1}$ addresses are accessible, see Fig.~\ref{fig:repair_as_you_go}.

We first partition the two sides of the QRAM into the ``repairable" and ``spare" QRAMs. We take the repairable QRAM to be that with the most functioning addresses, and the spare as the other side that will supply repair addresses \footnote{Other schemes are possible, where the QRAM is subdivided further into e.g. quarters. Faulty addresses in the two repairable quarters are then rerouted to spare addresses in the spare quarters. We focus on the simplest case of dividing the QRAM in half for simplicity, but optimizations along this route are nevertheless possible.}. Thus, faulty addresses in the repairable QRAM are assigned to available addresses in the spare QRAM. In the following, we speak of assigning routers to one another rather than addresses: this is because addresses associated with the same router can always be assigned together. Moreover, because the assignment is done layer by layer, it is more natural to speak of assigning routers to one another at each layer.

\SetCommentSty{textsl}
\begin{algorithm}
\caption{\label{alg:as_you_go} The \texttt{IterativeRepair} algorithm repairs the QRAM layer by layer. It utilizes an auxiliary function \texttt{GetMapping} which takes as input lists of faulty and available routers and outputs a repair mapping between them (this could for instance be the \texttt{FlagQubitMinimization} algorithm described in Sec.~\ref{sec:flag_min}).}
\SetKwFunction{func}{IterativeRepair}
\SetKwProg{Fn}{Function}{:}{}
\Fn{\func{r}}{
    \KwIn{router $r$ which is the root node of a faulty QRAM}
    \KwOut{OverallMapping, assignment of faulty routers to available routers and LayerMappings, intermediate repair at each layer}

    \BlankLine
    PrevMapping$\gets\emptyset$\;
    LayerMappings$\gets\emptyset$\;
    $d_t\gets$ depth of QRAM tree rooted at $r$\;
    $d_{rr} \gets $ largest depth QRAM available from \texttt{RelabelRepair}\;
    $c_{r}\gets$ child of $r$ corresponding to ``repairable" tree\;
    $c_{s}\gets$ child of $r$ corresponding to ``spare" tree\;
    \For{$d_i\in\{d_{rr}+1,\ldots, d_t\}$}{
        $f_r\gets$ faulty routers at depth $d_i$ in $c_{r}$\;
        $a_r\gets$ available routers at depth $d_i$ in $c_{s}$\;
        $u_r\gets$ faulty routers at depth $d_i$ in $c_{s}$ \;
        $r_r \gets$ all routers in $u_r$ whose parents are used in PrevMapping \tcp*[r]{routers previously assigned to repair faulty routers, themselves now needing repair}
        $p_r \gets$ all routers in $f_r$ whose parents were assigned to the parents of routers in $r_r$\tcp*[r]{need to track reassignments}
        append $r_r$ to $f_r$ \;
        remove from $f_r, a_{r}$ all routers whose parents appear in PrevMapping\tcp*[r]{either get assignment for free or reassignment covered by $r_r$}
        NewRepair $\gets$ \texttt{GetMapping($f_{r},a_{r}$)}\;
        OverallMapping $\gets$ assignments from NewRepair, where the routers in $p_r$ are assigned to the routers in $a_{r}$ where the respective routers in $r_r$ are mapped\;
        append to OverallMapping remaining valid assignments from PrevMapping\;
        append OverallMapping to LayerMappings\;
        PrevMapping $\gets$ OverallMapping
    }
    
    \KwRet{{\rm NewMapping, LayerMappings}}
}   
\end{algorithm}

We assume that we have access to a $k$-bit QRAM found using \texttt{RelabelRepair} where $k<n-1$ (the case $k=n-1$ is exactly the case where \texttt{RelabelRepair} alone succeeds). $k$ is of course at least 2 given that the top router and its two children are non-faulty by assumption. The first step of the iterative procedure is to construct a $k+1$-bit QRAM. We do this by first classically assigning faulty routers to repair routers at layer $k+2$, see Fig.~\ref{fig:repair_as_you_go}. (We put aside for now the question of how to best perform the assignment of faulty routers to repair routers, which we return to in Sec.~\ref{sec:flag}). We begin at $k+2$ because the user does not have access to the state of the most significant address qubit, which is only activated to access repair routers. That is to say, we specify the location of faulty routers at level $k+2$ using $k$ bits specified by the user in addition to the most significant address qubit whose state is determined by which side is the repairable QRAM.

Given an assignment, we perform the rerouting using ancilla flag qubits using one of two strategies described in detail in Sec.~\ref{sec:flag}: we outline only the main points here. In both cases, the flag qubits are sent in as bus qubits to the $k$-bit QRAM. Importantly, this strategy assumes that we can access classical data even in the {\it interior} of the QRAM tree, and not just at the leaves. This functionality is supported by QRAM hardware proposals, such as those in Ref.~\cite{Weiss2024}.

\begin{figure}
    \centering
    \includegraphics[width=\columnwidth]{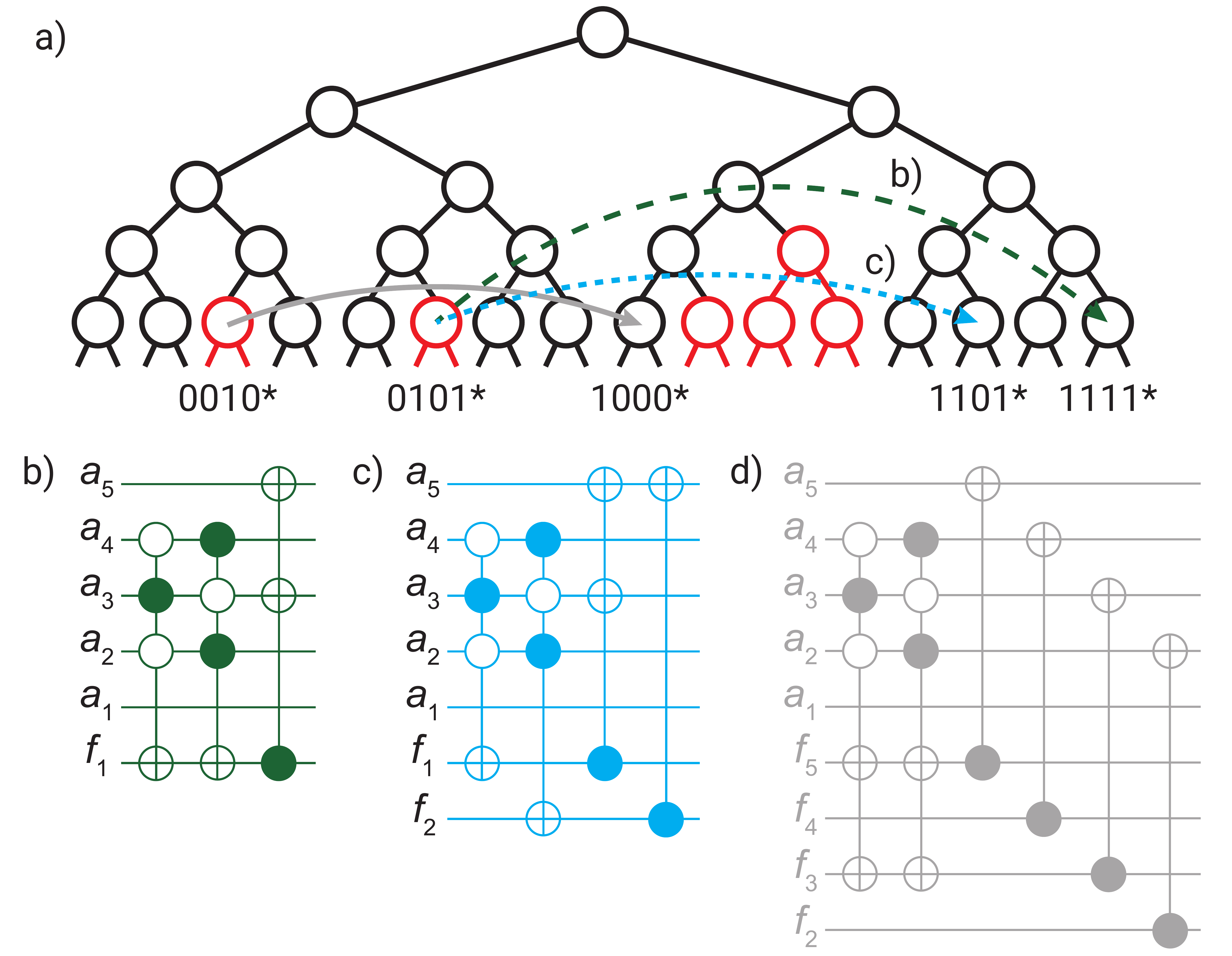}
    \caption{\label{fig:flag_qubit_min} The number of necessary flag qubits can be minimized by considering the bit-flip patterns needed to map faulty routers to available ones. (a) For the given faulty-router configuration, we consider two possible mappings. (b) For the router assignment indicated in (a) by the gray and green lines, we flip $a_{5}$ and $a_{3}$ for both faulty routers. Thus, only a single flag qubit is necessary. (c) For the router assignment indicated in (a) by the gray and blue lines, the bit-flip pattern now differs for the two repaired routers. This necessitates utilizing two flag qubits, which serve as bus qubits for two separate QRAM queries. (d) Using the flag-qubit mask technique, the mapping in (b) requires four flag qubits.}
\end{figure}

In the first flag-qubit strategy, the flag qubits serve as a mask that is applied to the address qubits to perform the rerouting. For a $j$-bit QRAM we require $j$ flag qubits, where each flag qubit is responsible for conditionally flipping a specific address qubit, see Fig.~\ref{fig:repair_as_you_go}(c, f). In the second strategy, we attempt to minimize the number of flag qubits by modifying their role: we assign a specific {\it address bit-flip pattern} to each flag qubit. That is, if two faulty addresses are rerouted using the same pattern of bit flips on the address qubits, we only need a single flag qubit to perform the rerouting, see Fig.~\ref{fig:flag_qubit_min}.

\begin{figure*}
    \centering
    \includegraphics[width=\linewidth]{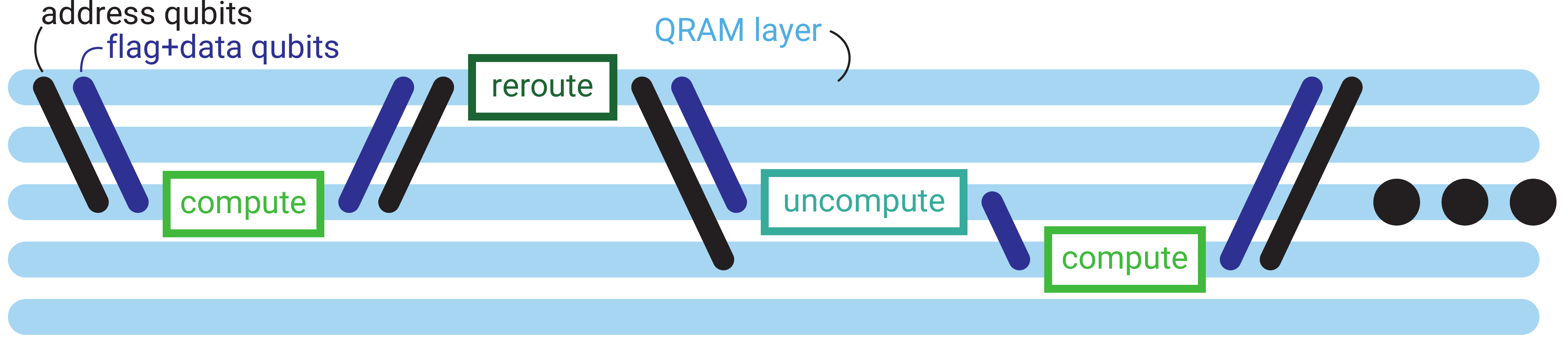}
    \caption{\label{fig:address_bus_route_schematic} To recursively repair the QRAM, address and bus qubits are routed in and out of the QRAM layer by layer.
    We first route the address and bus qubits in to the smallest available QRAM, such as the 2-bit QRAM shown in Fig.~\ref{fig:repair_as_you_go}(a, b). We then ``compute," where the flag qubits are activated according to the address qubits that need to be rerouted. To perform the ``reroute" operation, the address and bus qubits need to be first routed out of the tree. Only then can the CNOT operations in, e.g., Fig.~\ref{fig:repair_as_you_go}(c, f) be performed. The address qubits can now be routed one layer deeper in the tree. To disentangle the flag qubits from the address qubits they just rerouted, we then ``uncompute" as in Fig~\ref{fig:repair_as_you_go}(c, f). Indeed, this step can be combined with the next round of ``compute." The algorithm continues (indicated by the ellipses) until all layers have been repaired. 
    }
\end{figure*}

Once the flag qubits have been activated according to one of the two above strategies, they are unloaded from the QRAM tree to perform the rerouting on the address qubits, see Fig.~\ref{fig:address_bus_route_schematic}. Now, all queries that would have arrived at inaccessible routers at level $k+2$ instead are routed to accessible ones at the same level on the spare side of the tree. The final step of the repair at this level consists of uncomputing the flag qubits, see Fig.~\ref{fig:repair_as_you_go}. Uncomputing disentangles the flag qubits from the addresses and has the additional advantage of allowing flag qubits to be reused for repair of later layers. This step includes loading the repaired addresses back into the tree, followed by the bus qubits, see Fig.~\ref{fig:address_bus_route_schematic}. Importantly, the uncompute operations should be performed at the repaired address location, as opposed to the original address location, see Fig.~\ref{fig:repair_as_you_go}(c, f).

The address qubits have now been reloaded into the tree after performing the previous uncompute step, see Fig.~\ref{fig:address_bus_route_schematic}. Because of the rerouting, they can now be loaded one layer further down the tree than in the previous step. In this way, we now have access to a functioning $k+1$-bit QRAM, see Fig.~\ref{fig:repair_as_you_go}(d).
This QRAM can now be utilized to repair layer $k+3$ of the tree, yielding a functioning $k+2$-bit QRAM. As in the previous step, the address and bus qubits must then be unloaded from the tree to perform the rerouting, see Fig.~\ref{fig:address_bus_route_schematic}. They are then loaded back in to perform the uncompute step. This procedure continues until an $n-1$-bit QRAM has been obtained. 

We track the repaired routers from all previous steps, and at each step reassign faulty routers on the spare side that would otherwise be routed to based on previous repair steps, see Fig.~\ref{fig:repair_as_you_go}(d). Pseudocode for the \texttt{IterativeRepair} scheme outputting the repair mapping at each layer is given in Algorithm~\ref{alg:as_you_go}.

The final step is actually performing the QRAM query of interest.
The classical data that would have been associated with faulty address locations are now associated with the corresponding repaired addresses, and the query of the $n-1$-bit QRAM can proceed. 

In performing \texttt{IterativeRepair}, we pay a cost in terms of the time complexity of a QRAM query. In a standard query of a $j$-bit QRAM (without repair), the time complexity is $O(j + b)$, where $b$ is the number of bus qubits \cite{Chen2023}. The time cost is only linear in the number of address and bus qubits because both can be pipelined into the tree \cite{Chen2023, Xu2023, Jaques2023}. We show in Sec.~\ref{sec:flag} that the number of bus qubits does not exceed the number of address qubits at each layer during \texttt{IterativeRepair}, thus we take the time cost of repairing layer $j$ of the QRAM to be $O(j)$.
Repairing each layer of the QRAM from $k+2$ to $n$ thus increases the overall time complexity to $O(n^2)$, as we must repair the layers serially, see Fig.~\ref{fig:address_bus_route_schematic}. While the repair procedure does incur this increase in the time complexity, it is still only poly-logarithmic in the size of the classical memory $N$.

\section{Classical Reassignment}
\label{sec:flag}
We now address the question of the classical assignment of faulty routers to available routers deferred from Sec.~\ref{sec:iterative}. The assignment strategy depends on the role that we assign to the ancilla flag qubits that perform the rerouting. We discuss two possibilities here, though more are possible. The first is using the flag qubits as a mask, with one flag qubit for every address qubit. The second is using the flag qubits to each effect a specific bit-flip pattern on some subset of the address qubits. This second strategy reduces the number of required flag qubits as compared to the mask strategy.

\subsection{Overview of the Flag-qubit mask algorithm}

With the flag-qubit mask strategy, the flag qubits apply a mask of CNOTs to the address qubits, see Fig.~\ref{fig:repair_as_you_go}(c). Each flag qubit is assigned to conditionally flip a specific address qubit. Clearly, then, for a $j$-bit QRAM, we require $j$ flag qubits. A specific flag qubit is activated if two conditions are satisfied: (i) there is a query to a faulty router, and (ii) in the assignment of this faulty router to an available one, the address qubit assigned to this flag qubit is flipped. 

For the flag-qubit mask technique, the actual repair assignment of faulty addresses to available addresses is relatively unimportant. This is because the number of flag qubits as well as the bit-flip pattern applied to the address qubits is fixed, see Fig.~\ref{fig:repair_as_you_go}(c, f) (with the caveat that some CNOTs that will never be activated can be eliminated). Different assignments change the operations applied to the flag qubits, which only modify the classical data, see Fig.~\ref{fig:repair_as_you_go}(b, e). Thus, any random assignment of faulty routers to available routers suffices.

\subsection{Overview of the \texttt{FlagQubitMinimization} algorithm}
\label{sec:flag_min}

\SetCommentSty{textsl}
\begin{algorithm}[t]
\caption{\label{alg:flag_qubit_min} The \texttt{FlagQubitMinimization} algorithm seeks to limit the number of unique bit-flip patterns necessary to repair a list of faulty routers. }
\SetKwFunction{func}{FlagQubitMinimization}
\SetKwProg{Fn}{Function}{:}{}
\Fn{\func{{\rm FaultyRouterList}, {\rm AvailableRouterList}}}{
    \KwIn{FaultyRouterList and AvailableRouterList which specify the faulty and available routers, respectively}
    \KwOut{$RA$: Assignment of faulty routers to available routers}

    \BlankLine
    $RA\gets\emptyset$\;
    $GS\gets\emptyset$\;
    $M \gets $ bit-flip pattern matrix between routers in FaultyRouterList and AvailableRouterList\;
    \While{\rm Not all faulty routers have been assigned}{
    $p\gets$ bit-flip pattern that together with $GS$ generates a power set of bit-flip patterns that assigns the most unique faulty routers to the most unique available routers\;
    update $RA$ with corresponding assignments\;
    append $p$ to $GS$\;
    delete rows and columns of $M$ corresponding to faulty and available routers just assigned
    }
    \KwRet{$RA$}
}   
\end{algorithm}

By modifying the role played by the flag qubits, it is possible to optimize the number of flag qubits that are necessary to perform the rerouting. Instead of conditionally flipping a single address qubit, each flag qubit is instead now responsible for a specific bit-flip pattern, see Fig.~\ref{fig:flag_qubit_min}. Thus, two (or more) faulty routers repaired by the same bit-flip pattern are rerouted with the same flag qubit. Moreover, bit-flip patterns can be combined by activating multiple flag qubits. Thus, it is not just the chosen bit-flip patterns that can be used to reroute faulty addresses, but all bit-flip patterns in the power set of all combinations (bitwise addition modulo 2) of the chosen patterns. We refer to the set of chosen bit-flip patterns as the {\it generating set} and the set of all possible bit-flip patterns generated by this set as the {\it power set}.

The number of required flag qubits thus depends crucially on the chosen assignment of faulty routers to available ones. It could range from one flag qubit (if all faulty routers are repaired with the same bit-flip pattern, assuming there is at least one router needing repair) to $n-3$. The upper limit is $n-3$ because the two addresses accessed by the same router are always repaired together, we only repair routers in one half of the tree and at most half of those routers require repair (since that half of the tree is declared the ``repairable" side). So the number of faulty routers needing repair is limited to $2^{n-3}$, and $2^{n-3}$ bit-flip patterns can be generated by $n-3$ linearly independent bit-flip patterns. 

\sloppy The most naive method of finding the optimal assignment of faulty routers to available ones is checking all possible assignments. With $O[\epsilon\log_{2}(N)N]$ faulty routers and $O(N)$ available routers \footnote{This assumes $\epsilon \log_{2}(N)\ll 1$, which is not valid for the largest considered $\epsilon, n$ of e.g. 0.08, 13, respectively. In such cases the vast majority of QRAM instances are unrepairable. We then expect the small fraction of instances we are able to attempt repair on to have roughly $O(N)$ faulty and available routers.}, the time-complexity of this brute-force check is $O(N!)$. Such an algorithm is clearly not scalable.

We instead present a greedy algorithm \texttt{FlagQubitMinimization} for performing this assignment that attempts to minimize the number of flag qubits. The algorithm is presented as pseudocode in Algorithm~\ref{alg:flag_qubit_min}. The intuition for this algorithm is that at each step, we choose the bit-flip pattern that, together with the generating set, repairs the most faulty routers. 
We proceed by first computing the bit-flip pattern matrix $M$ between faulty and available routers. In the first iteration, we extract the most frequently occurring bit-flip pattern $p$ in $M$ and assign the corresponding pairs of faulty and available routers mapped to each other by $p$. These routers are then removed from the pool of faulty and available routers. 
The bit-flip pattern $p$ is now added to the generating set $GS$, which prior to the addition of $p$ contained only the zero bit-flip pattern. 

In the next and later loop iterations, we choose the bit-flip pattern $p'$ that together with the generating set $GS$ generates a power set of bit-flip patterns that can assign the most unique faulty routers to unique available routers. The key here is that all bit-flip patterns that are newly available by inclusion of $p'$ in the generating set can be used to reroute faulty routers. For example, in the second iteration, the newly available bit-flip patterns would be $p'$ and $p\oplus p'$.
We continue in this way until all faulty routers have been repaired.

\section{Evaluation}
\label{sec:evaluation}

\subsection{Experimental Setup}
Monte-Carlo simulations are implemented on a high-performance-computing cluster on multiple individual nodes with up to 8 cores per node to parallelize different experimental QRAM realizations. The largest simulations require 1.3 GB of RAM. For smaller depth QRAMs with $n=3,4,5$ we realize $1,000,000$ shots to generate sufficient statistics. For larger depth QRAMs $n=6,\ldots,13$ we generate $100,000$ shots. For each QRAM instance of depth $n$ we run the \texttt{RelabelRepair} algorithm for $3\leq m \leq n$, recording whether the algorithm succeeds. We then run the full \texttt{IterativeRepair} algorithm. We also separately run the \texttt{FlagQubitMinimization} algorithm on only the final layer of the QRAM to independently evaluate the performance of this algorithm. In this case, we simply attempt to assign all faulty routers at the lowest layer to available routers at the same layer.

\subsection{Repair Simulations}

\begin{figure}
    \centering
    \includegraphics[width=\linewidth]{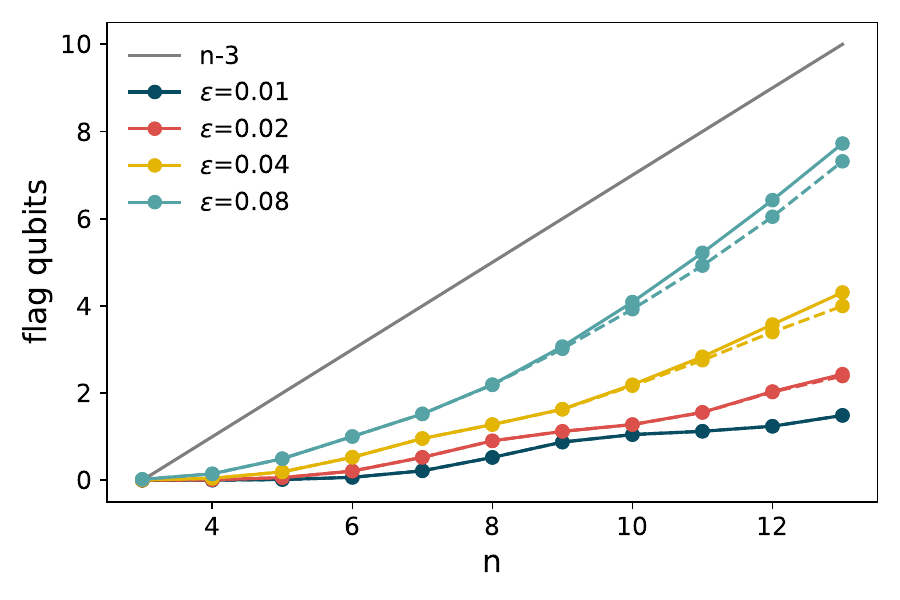}
    \caption{
    \label{fig:flag} Among QRAM instances that have at least $2^{n-1}$ available addresses, we run \texttt{FlagQubitMinimization} on the lowest layer of the QRAM (solid) as well as the full \texttt{IterativeRepair} algorithm (dashed) and compute the average number of required flag qubits. Remarkably, even in cases where tens or hundreds of routers are faulty, only a few flag-qubits are required to perform the rerouting.
    The gray line shows the expected ceiling at $n-3$ for the number of flag-qubits. The lines are a guide to the eye. 
    }
\end{figure}
Despite tens or hundreds of faulty addresses, in many cases only a handful of flag qubits are required for repair, see Fig.~\ref{fig:flag}. 
For instance, with $n=13$ and $\epsilon=0.01$, we expect $(1-(1-0.01)^{11})2^{13}=857$ faulty addresses, thus 429 faulty routers. About half of these routers are on the repairable side of the tree, thus 214 routers need repair. Nevertheless, only 1.5 flag qubits are required on average for both \texttt{IterativeRepair} and \texttt{FlagQubitMinimization}, see Fig.~\ref{fig:flag}. This represents a reduction in the number of flag qubits required as compared to the naive flag-qubit mask strategy by a factor of 13/1.5 = 8.7. 
We note importantly that flag qubits can be reused in repairing one layer to the next in \texttt{IterativeRepair}. Thus, in counting the number of flag qubits used by \texttt{IterativeRepair}, we take the maximum over the number of flag qubits used at each layer (typically the final layer). Moreover, the average number of flag qubits required by \texttt{IterativeRepair} is generally lower than that of \texttt{FlagQubitMinimization} applied only to the bottom layer, see Fig.~\ref{fig:flag}. This is due to the reduction in the effective number of faulty routers at the final layer of the tree in the former case, from repairs executed at higher levels of the tree during the iterative procedure.

\begin{figure}
    \centering
    \includegraphics[width=\linewidth]{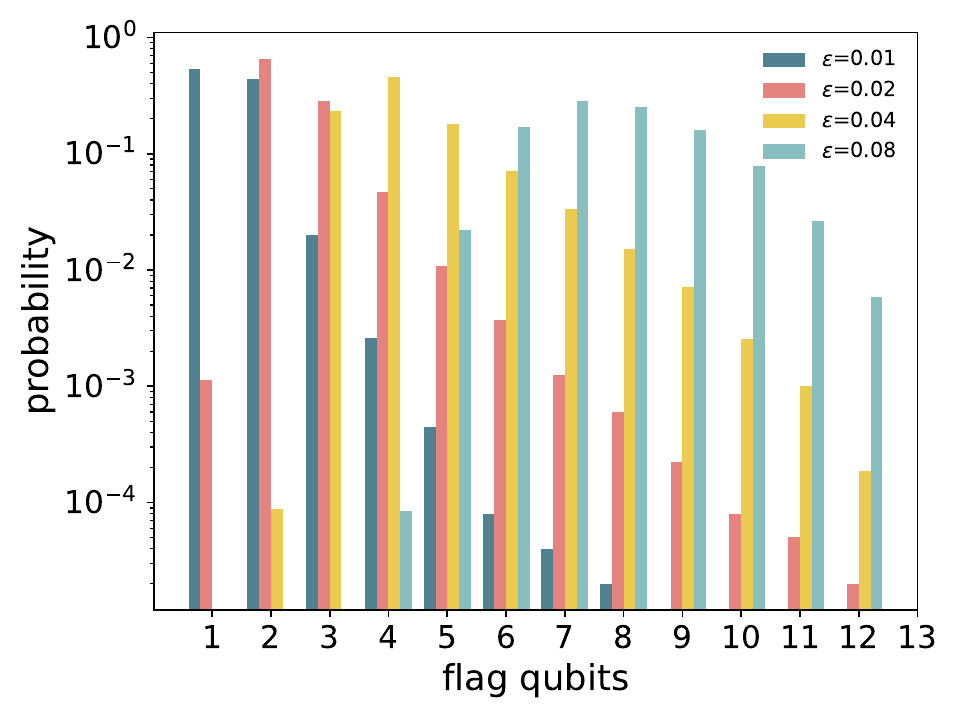}
    \caption{
    \label{fig:num_flag_hist} The distribution of the number of required flag qubits for $n=13$ from the \texttt{FlagQubitMinimization} algorithm. We observe that the distribution is skewed, and in some cases the greedy algorithm requires up to $n-1$ flag qubits.}
\end{figure}

It is important to note that because the \texttt{FlagQubitMinimization} algorithm is greedy and does not perform a global optimization, it can sometimes yield a suboptimal solution. For instance, while in principle we should never require more than $n-3$ flag qubits to perform a repair, the algorithm can in rare cases require up to $n-1$ flag qubits, see Fig.~\ref{fig:num_flag_hist}. This generally occurs when the pool of faulty routers is nearly equal in size to the pool of available routers, restricting the number of bit-flip patterns that can be used for repair. The algorithm can then ``back itself into a corner," only assigning one or a few faulty routers at a time before eventually placing $n-1$ flag qubits in the generating set. In such cases, the generating set can be discarded in favor of the linearly-independent set of at most $n-3$ bit-flip patterns that generates the required bit-flip patterns. (Indeed, any random assignment can be chosen, requiring at most $n-3$ bit-flip patterns).

\begin{figure}
    \centering
    \includegraphics[width=\linewidth]{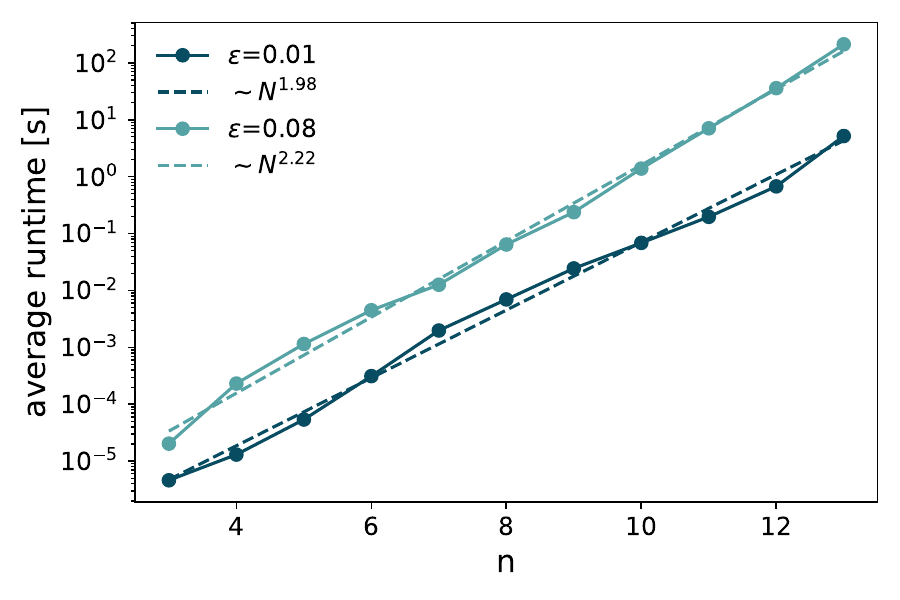}
    \caption{\label{fig:avg_runtime} Average runtime of \texttt{FlagQubitMinimization} to repair the lowest layer of the QRAM. The dashed lines are the best fit to $a N^b$, indicating that the runtime increases approximately quadratically with the physical memory size $N$.}
\end{figure}

We generally expect the runtime of \texttt{FlagQubitMinimization} applied to the final layer (in the absence of parallelization) to scale asymptotically as $O[\epsilon\log_{2}(N)N^3]$ due to two contributions. The first is the scan over $O(N)$ possible bit-flip patterns. The second is scanning over the $O[\epsilon\log_{2}(N)N^2]$ elements in the bit-flip-pattern matrix between faulty and available routers. (There is additionally the loop over the number of flag qubits utilized, however the faulty and available router lists are decreasing in size after every iteration. Thus, this contribution should be subdominant.) The numerical runtime of \texttt{FlagQubitMinimization} is observed to scale approximately quadratically with $N$ for a given $\epsilon$, see Fig.~\ref{fig:avg_runtime}. We have observed (not shown) that the runtime of \texttt{IterativeRepair} is generally approximately equal to that of \texttt{FlagQubitMinimization} run on the final QRAM layer, indicating that the final step of \texttt{IterativeRepair} is rate limiting.  We attribute the difference between the observed $\sim O(N^2)$ and predicted $O[\log_{2}(N)N^3]$ scaling (for a given $\epsilon$) to finite-size effects of the numerical implementation.

\section{Discussion and Related Works}
\label{sec:discussion}

Previous work by Kim {\it et al.} considered the issue of faulty memory cells in a bucket-brigade QRAM architecture \cite{Kim2023}. In a QRAM used for accessing classical data, such quantum memory cells do not exist; the classical data is stored only in the classical controller before it is copied into the state of the bus (after it has been routed to the bottom of the tree). One can view the work of Kim {\it et al.} in this context as instead applying to the quantum routers at the bottom layer of the tree, each responsible for accessing two address locations. Our work goes beyond this by considering faulty routers in the interior of the tree, which have an outsize effect on the number of resulting faulty addresses as compared to faulty routers only at the bottom layer. 

Kim {\it et al.} propose to remedy the issue of faulty quantum memory cells with redundant quantum memory cells. Here, we instead consider an architecture where an $n$-bit, possibly faulty QRAM has been constructed, and we want to extract a functioning $n-1$-bit QRAM. This can be viewed as providing twice as many physical qubits as would normally be required. In our architecture, we retain the usual QRAM binary-tree structure \cite{Giovannetti2008, Giovannetti2008architectures, Hann2021}.

It is worthwhile in future work to consider alternative QRAM architectures that go beyond binary trees. The binary-tree architecture ensures that the memory is random access and that there is a unique path from root to each address. However, as discussed in this work, it is susceptible to router failures that render addresses below such routers inaccessible. Faulty routers occurring deep in the tree can be mitigated, as such failures only affect a small fraction of the total number of addresses. However, router failures higher up in the tree are much more problematic, rendering large swaths of the tree inaccessible. One could consider a hybrid architecture with redundant routers at higher layers to combat this issue.

\section{Conclusion}
\label{sec:conclusion}

As we enter the era where quantum devices move from tens to hundreds or even thousands of qubits, robust techniques are required for mitigating inevitable fabrication defects. Here, we present novel algorithms for routing around faulty routers to recover an $n-1$-bit QRAM from a physical $n$-bit QRAM.

Our \texttt{RelabelRepair} algorithm works by relabeling address locations and assigning certain routers to route in only one direction. This algorithm does not always succeed, as one-way routers inevitably cause some available routers to not be utilized. 

In such cases, we deploy the \texttt{IterativeRepair} algorithm, which recursively constructs larger-depth QRAMs building upon smaller depth devices. Having obtained a classical mapping of faulty routers to available ones at each layer, we quantum-mechanically reroute queries from faulty routers to available ones without perturbing queries to non-faulty routers. The rerouting is effected by ancilla flag qubits, which can be utilized in one of two ways. The first, more pedagogical method, is applying the flag qubits as a mask onto the address qubits. This technique requires the same number of flag qubits as address qubits. The second technique optimizes the number of flag qubits by assigning a single bit-flip pattern to each flag qubit. This idea motivates assigning faulty routers to available ones using the minimal number of bit-flip patterns. We present a classical algorithm for obtaining an assignment that minimizes the number of flag qubits used at each layer. While the obtained classical assignments can be reused from one query to the next, the iterative rerouting using ancilla flag qubits must be repeated for every new query, as described in Fig.~\ref{fig:address_bus_route_schematic}. While this procedure does incur an increase in the time complexity of a QRAM query, from $O(n)$ to $O(n^2)$, it remains poly-logarithmic in the size of the memory $N$.

\section{Acknowledgments}

We acknowledge useful conversations with Nathan Wiebe and Gaurav Mahajan. We thank the Yale Center for Research Computing, specifically Tom Langford, for guidance and assistance in computations run on the Grace cluster. D.~K.~W, S.~P. and S.~M.~G acknowledge support from the Air Force Office of Scientific Research under award number FA9550-21-1-0209. S.~X. and Y.~D. were supported by the National Science Foundation under award CCF-2312754. The U.S. Government is authorized to reproduce and distribute reprints for Governmental purposes notwithstanding any copyright notation thereon. The views and conclusions contained herein are those of the authors and should not be interpreted as necessarily representing the official policies or endorsements, either expressed or implied, of the Air Force Office of Scientific Research or the U.S. Government.

External interest disclosure: S.P., S.M.G., and Y.D.\ receive consulting fees from Quantum Circuits, Inc., and S.M.G.\ is an equity holder in Quantum Circuits, Inc.

We have made the code utilized in this work publicly available at Ref.~\cite{github}.

\bibliographystyle{IEEEtran}
\bibliography{bib}

\end{document}